\documentclass[a4paper,11pt]{article}
\usepackage{pos}
 \usepackage[nice]{nicefrac}
\usepackage[utf8]{inputenc}
\usepackage{graphicx}
\usepackage{amsmath}
\usepackage{amsfonts}
\usepackage{physics}
\usepackage{xspace}
\usepackage{bm}
\usepackage{mathrsfs}
\usepackage{nicefrac}
\usepackage{multirow}
\usepackage[format=plain,justification=RaggedRight,labelsep=endash,font=small,labelfont=sc]{caption}

\usepackage[export]{adjustbox}

\definecolor {darkgreen}{rgb}{0.2,0.7,0.2}
\newcommand{\pdagger}{{\phantom{\dagger}}}
 \newcommand{\psbar}{\bar{\psi}}

\def\be{\begin{equation}}
	\def\ee{\end{equation}}
\newcommand{\bel}[1]{\begin{eqnarray}\label{#1}}
	\newcommand{\eel}{\end{eqnarray}}
\def\barr{\begin{array}}
	\def\earr{\end{array}}
\def\beq{\begin{eqnarray}}
	\def\eeq{\end{eqnarray}}
\def\bfig{\begin{figure}}
	\def\efig{\end{figure}}

\def\n0{n_{(0)}}
\def\e0{\varepsilon_{(0)}}
\def\P0{P_{(0)}}
\title{Mathematical expressions for quantum fluctuations of energy for different energy-momentum tensors}

\author*{Rajeev Singh}
\affiliation{Institute of Nuclear Physics Polish Academy of Sciences, PL-31-342 Krak\'ow, Poland}

\emailAdd{rajeev.singh@ifj.edu.pl}

\abstract{Expressions for the quantum fluctuations of energy density have been derived for the subsystems consisting of hot relativistic gas of particles with spin-$\frac{1}{2}$ and mass $m$. Our expressions for the fluctuation depend on the form of energy-momentum tensor which in turn depend on the choice of pseudo-gauge. These results suggest that quantum fluctuations of energy should be considered seriously in the case of the very small thermodynamic systems.}

\FullConference{%
  The Ninth Annual Conference on Large Hadron Collider Physics-LHCP2021\\
  7-12 June, 2021\\
  online}

\begin{document}
\maketitle
\section{Introduction}
\label{sec:intro}
Quantum and statistical fluctuation intrinsic to any many-body system~\cite{Huang:1987asp} has an important role as they contain crucial information about the possible phase transitions~\cite{Huang:1987asp,Smoluchowski,PhysRevLett.85.2076,Gross:1980br,Haussler:2008gx,Herzog:2016kno,Vovchenko:2016dix,Steinheimer:2016cir,Lohr-Robles:2021ihl,Bai:2021wrh,Fortunato:2021fby,Li:2021qcb,Yang:2021hmg,Sami:2021ufn,deBrito:2021pmw}, dissipative phenomena~\cite{Kubo1,Berges:1998rc,Halasz:1998qr,Stephanov:1998dy,Stephanov:1999zu,Hatta:2002sj,Son:2004iv,Stephanov:2008qz,Berdnikov:1999ph,CaronHuot:2011dr,Kitazawa:2013bta,Bluhm:2020mpc,Goswami:2020yez,Fu:2021wyc,Pradeep:2021opj,Goswami:2021opr,Bollweg:2021vqf,Schmidt:2021pey,Guo:2021qtf}, and the formation of the large scale structures in the universe~\cite{Lifshitz:1963ps,PhysRevLett.49.1110,Choudhury:2013woa,Choudhury:2016cso,Choudhury:2016pfr,Gelis:2016upa,Gelis:2019vzt,Choudhury:2021tuu,Graef:2021tid}. 
Following the footsteps of our previous works~\cite{Das:2020ddr,Das:2021aar,Das:2021rck,Singh:2021dbu,Singh:2021xiv,Singh:2021tih}, we analyze and study the pseudo-gauge dependence on the quantum fluctuations of energy, which means, using the pseudo-gauge transformation we can choose different form of the energy-momentum tensor for the description of the system.
Any energy-momentum tensor $\hat{T}^{\mu\nu}$ satisfying the conservation equation $\partial_\mu \hat{T}^{\mu\nu}=0$ can be used to construct a new conserved energy-momentum tensor as
~\cite{Chen:2018cts,HEHL197655,Speranza:2020ilk}
\begin{equation}
\hat{T}^{\prime \,\mu\nu} = \hat{T}^{\mu\nu} + \partial_\lambda \hat{A}^{\nu\mu \lambda} \quad \text{with} \quad \hat{A}^{\nu\mu \lambda} = - \hat{A}^{\nu\lambda \mu}.
\label{eq:PG}
\end{equation}
For the current analysis, we consider a system of spin-$\frac{1}{2}$ particles and study the effects of different pseudo-gauges. We choose three different forms of energy-momentum tensor, such as canonical form from the Noether theorem~\cite{BELINFANTE1939887,BELINFANTE1940449,Rosenfeld1940}, the de Groot-van Leeuwen-van  Weert (GLW) form~\cite{DeGroot:1980dk}, and the Hilgevoord-Wouthuysen (HW) form~\cite{HILGEVOORD19631,HILGEVOORD19651002}. These forms are currently being discussed widely in the context of heavy-ion collisions for the study of spin polarization~\cite{Florkowski:2018fap,Speranza:2020ilk}, as the pseudo-gauge choices can also be applied for the spin tensor $\hat{S}^{\lambda, \mu\nu}$ which is a part of the total angular momentum tensor $\hat{J}^{\lambda, \mu\nu} = \hat{L}^{\lambda, \mu\nu}+\hat{S}^{\lambda, \mu\nu}$~\cite{HEHL197655,Garbiso:2020puw,Singh:2020efc,Speranza:2020ilk,Leader:2013jra,Gallegos:2021bzp,Singh:2021liw}.
We calculate quantum fluctuation of the $\hat{T}^{00}$ component of the energy-momentum tensor and find that even though $\hat{T}^{00}$ depends on pseudo-gauge, its thermal average value is independent of it. In addition, we see that for small size subsystems the fluctuations are pseudo-gauge dependent but become independent if the size of the system is large. This analysis might be useful for the understanding of the concept of energy-density in the context of relativistic heavy-ion collisions~\cite{Dusling:2010rm,Epelbaum:2013waa,Jaiswal:2016hex,Florkowski:2017olj,Romatschke:2017ejr,Giacalone:2019kgg,Bhalerao:2020ulk,Becattini:2020ngo,Bhadury:2021oat,Sarwar:2021csp,Dore:2021xqq,Karthein:2021cmb}. Our conclusion is that the quantum fluctuations of energy-density has no physical significance in small size subsystems and specific pseudo-gauge must be chosen in order to describe the system. 
\section{Basic definitions}
\label{sec:basic}
As like our previous studies~\cite{Das:2020ddr,Das:2021aar,Das:2021rck}, here we assume a subsystem
$S_a$
inside a larger thermodynamic system $S_V$ 
consisting of spin-$\frac{1}{2}$ particles having mass $m$
with no conserved charges. The volume $V$ of the system $S_V$
is large enough to perform integrals over particle momentum. 
\footnote{Metric $g_{\mu\nu} = \hbox{diag}(+1,-1,-1,-1)$ is used. Three-vectors are shown in bold font and a dot is used to denote the scalar product of both four- and three-vectors, i.e., $a^\mu b_\mu = a \cdot b = a^0 b^0 - \boldsymbol{a} \cdot  \boldsymbol{b}$.  } 
We describe our system by a spin-$\frac{1}{2}$ field in thermal equilibrium where the field operator is~\cite{Tinti:2020gyh}
\begin{align}
\psi(t,\boldsymbol{x})=&\sum_r\int\frac{d^3k}{(2\pi)^3\sqrt{2\omega_{\boldsymbol{ k}}}}\Big(U_r^{\pdagger}(\boldsymbol{k})a_r^{\pdagger}(\boldsymbol{k})e^{-i k \cdot x} +V_r^{\pdagger}(\boldsymbol{k})b_r^{\dagger}(\boldsymbol{k})e^{i k \cdot x} \Big),
\label{equ1ver1}
\end{align}
with $a_r^{\pdagger}(\boldsymbol{k})$ and $b_r^{\dagger}(\boldsymbol{k})$ being the annihilation and creation operators for particles and antiparticles, respectively, satisfying the anti-commutation relations, $\{a_r^{\pdagger}(\boldsymbol{k}),a_s^{\dagger}(\boldsymbol{k}^{\prime})\} =(2\pi)^3\delta_{rs} \delta^{(3)}(\boldsymbol{k}-\boldsymbol{k}^{\prime})$ and
$ \{b_r^{\pdagger}(\boldsymbol{k}),b_s^{\dagger}(\boldsymbol{k}^{\prime})\} =(2\pi)^3\delta_{rs} \delta^{(3)}(\boldsymbol{k}-\boldsymbol{k}^{\prime})$. The index $r$ is the polarization degree of freedom, and $U_r^{\pdagger}(\boldsymbol{k})$ and $V_r^{\pdagger}(\boldsymbol{k})$ are the Dirac spinors where $\omega_{\boldsymbol{k}}=\sqrt{\boldsymbol{k}^2+m^2}$ is the energy of a particle. 

The following expectation values are required to calculate thermal averages~\cite{CohenTannoudji:422962,Itzykson:1980rh,Evans:1996bha} of the energy-density operator $\hat{T}^{00}_a$
\begin{align}
& \langle a_r^{\dagger}({\boldsymbol{k}})a_s^{\pdagger}({\boldsymbol{k}}^{\prime})\rangle=(2\pi)^3\delta_{rs}\delta^{(3)}({\boldsymbol{k}}-{\boldsymbol{k}}^{\prime})f(\omega_{\boldsymbol{k}}),\label{equ2ver1}\\
& \langle a^{\dagger}_r(\boldsymbol{k})a^{\dagger}_s(\boldsymbol{k}^{\prime})a_{r^{\prime}}^{\pdagger}(\boldsymbol{p})a_{s^{\prime}}^{\pdagger}(\boldsymbol{p}^{\prime})\rangle =(2\pi)^6 \Big(\delta_{rs^{\prime}}\delta_{r^{\prime}s}\delta^{(3)}(\boldsymbol{k}-\boldsymbol{p}^{\prime})~\delta^{(3)}(\boldsymbol{k}^{\prime}-\boldsymbol{p})\nonumber\\
&-\delta_{rr^{\prime}}\delta_{ss^{\prime}}\delta^{(3)}({\boldsymbol{k}}-\boldsymbol{p})~\delta^{(3)}({\boldsymbol{k}}^{\prime}-\boldsymbol{p}^{\prime})\Big)f(\omega_{{\boldsymbol{k}}})f(\omega_{{\boldsymbol{k}}^{\prime}}).\label{equ4ver1}
\end{align}
where $f(\omega_{{\boldsymbol{k}}})$ is the Fermi--Dirac distribution function for particles.
$\hat{T}^{00}_a$ is an energy-density operator expressed below of the subsystem $S_a$ which is placed at the origin of coordinate system~\cite{Chen:2018cts}, and we use Gaussian profile to define our subsystem $S_a$ 
in order to have no sharp-boundary effects
\begin{align}
\hat{T}^{00}_a = \frac{1}{(a\sqrt{\pi})^3}\int d^3\boldsymbol{x}~\hat{T}^{00}(x)~\exp\left(-\frac{{\boldsymbol{x}}^2}{a^2}\right).
\label{equ6ver1}
\end{align}
Then, we calculate the variance ($\sigma^2$) and the normalized standard deviation ($\sigma_n$) using expressions below in order to find the fluctuation of the energy density of the subsystem $S_a$,
\beq
 \sigma^2(a,m,T) = \langle :\hat{T}^{00}_a: :\hat{T}^{00}_a: \rangle - \langle :\hat{T}^{00}_a :\rangle^2\,, \quad 
\sigma_n(a,m,T)= \frac{(\langle:\hat{T}^{00}_a::\hat{T}^{00}_a:\rangle- \langle :\hat{T}^{00}_a :\rangle^2)^{1/2}}{\langle :\hat{T}^{00}_a :\rangle}.
\label{sigma2}
\eeq
where $\langle :\hat{T}^{00}_a :\rangle$ is the thermal expectation value of $\hat{T}^{00}_a$ after doing normal ordering.
\section{Energy density fluctuation in different pseudo-gauges}
\label{section3}
\subsection{Canonical framework}
\label{sec:can}
The canonical form of energy-momentum tensor is given as~\cite{Tinti:2020gyh}
 \begin{align}
     \hat{T}^{\mu\nu}_{\text{Can}}=\frac{i}{2}\bar\psi\gamma^{\mu}\overleftrightarrow{\partial}^{\nu}\psi.
     \label{equ9ver1}
 \end{align}
where the thermal expectation value of $\hat{T}^{00}_{\text{Can},a}$ for the subsystem $S_a$
is calculated as
\begin{align}
\langle:\hat{T}^{00}_{\text{Can},a}:\rangle
& = 4\int\frac{d^3k}{(2\pi)^3}~\omega_{\boldsymbol{k}}~f(\omega_{\boldsymbol{k}}) \equiv\varepsilon_{\text{Can}}(T). 
\label{equ11ver1}
\end{align}
with factor $4$ representing the spin degeneracy ($g_s=(2s+1)$). Canonical energy-density $\varepsilon_{\text{Can}}(T)$, Eq.~\eqref{equ11ver1}, is independent of both time and the system size $a$ indicating the system's spatial uniformity.
Then we calculate the energy-density fluctuation for $\hat{T}^{\mu\nu}_{\text{Can}}$ as
\begin{align}
\sigma^2_{\text{Can}}(a,m,T) =  2\int dK ~dK^{\prime} f(\omega_{{\boldsymbol{k}}})(1-f(\omega_{{\boldsymbol{k}}^{\prime}}))\times \bigg[(\omega_{{\boldsymbol{k}}}+\omega_{{\boldsymbol{k}}^{\prime}})^2(\omega_{{\boldsymbol{k}}}\omega_{{\boldsymbol{k}}^{\prime}}+{\boldsymbol{k}}\cdot{\boldsymbol{k}}^{\prime}+m^2)e^{-\frac{a^2}{2}({\boldsymbol{k}}-{\boldsymbol{k}}^{\prime})^2}\nonumber\\
~~~~~~~~~~~-(\omega_{{\boldsymbol{k}}}-\omega_{{\boldsymbol{k}}^{\prime}})^2(\omega_{{\boldsymbol{k}}}\omega_{{\boldsymbol{k}}^{\prime}}+{\boldsymbol{k}}\cdot{\boldsymbol{k}}^{\prime}-m^2)e^{-\frac{a^2}{2}({\boldsymbol{k}}+{\boldsymbol{k}}^{\prime})^2}\bigg],
\label{equ12ver1}
\end{align}
where $dK \equiv d^3{{k}}/((2\pi)^{3} 2 \omega_{{\boldsymbol{k}}})$. In Eq. \ref{equ12ver1}, we neglect a temperature-independent term to remove all vacuum divergences~\cite{Das:2020ddr}.
%
\subsection{de~Groot-van~Leeuwen-van~Weert framework}
\label{sec:GLW}
The de~Groot-van~Leeuwen-van~Weert form of energy-momentum tensor is given as~\cite{DeGroot:1980dk}
\begin{align}
    \hat{T}^{\mu\nu}_{\text{GLW}}=\frac{1}{4m}\Big[-\bar{\psi}(\partial^{\mu}\partial^{\nu}\psi)+(\partial^{\mu}\bar{\psi})(\partial^{\nu}\psi) +(\partial^{\nu}\bar{\psi})(\partial^{\mu}\psi)-(\partial^{\mu}\partial^{\nu}\bar{\psi})\psi\Big].
    \label{equ14ver1}
\end{align}
In this case we obtain expressions for the thermal average and fluctuation as
    \begin{align}
&\langle:\hat{T}^{00}_{\text{GLW},a}:\rangle
= 4\int\frac{d^3k}{(2\pi)^3}~\omega_{\boldsymbol{k}}f(\omega_{\boldsymbol{k}}) \equiv\varepsilon_{\text{GLW}}(T)
\label{equ15ver1}\\
&\sigma^2_{\text{GLW}}(a,m,T) =  \frac{1}{2m^2}\int dK dK^{\prime} f(\omega_{{\boldsymbol{k}}})(1-f(\omega_{{\boldsymbol{k}}^{\prime}}))\bigg[(\omega_{{\boldsymbol{k}}}+\omega_{{\boldsymbol{k}}^{\prime}})^4\left(\omega_{{\boldsymbol{k}}}\omega_{{\boldsymbol{k}}^{\prime}}-{\boldsymbol{k}}\cdot{\boldsymbol{k}}^{\prime}+m^2\right)\nonumber\\
&~~~~~~~~~~~~~~~~~~~~\times e^{-\frac{a^2}{2}({\boldsymbol{k}}-{\boldsymbol{k}}^{\prime})^2} -(\omega_{{\boldsymbol{k}}}-\omega_{{\boldsymbol{k}}^{\prime}})^4 \left(\omega_{{\boldsymbol{k}}}\omega_{{\boldsymbol{k}}^{\prime}}-{\boldsymbol{k}}\cdot{\boldsymbol{k}}^{\prime}-m^2\right)e^{-\frac{a^2}{2}({\boldsymbol{k}}+{\boldsymbol{k}}^{\prime})^2}\bigg]
\label{equ16ver1}
\end{align}
respectively. We again discard a divergent term which is temperature-independent. We note that, even though the thermal averages for the canonical and GLW energy-momentum tensors are same, $\langle:\hat{T}^{00}_{\text{Can},a}:\rangle$ = $\langle:\hat{T}^{00}_{\text{GLW},a}:\rangle$, but their fluctuations are not, $\sigma^2_{\text{Can}}(a,m,T)$ $\neq$ $\sigma^2_{\text{GLW}}(a,m,T)$. 
\subsection{Hilgevoord-Wouthuysen framework}
\label{sec:HW}
The Hilgevoord-Wouthuysen form of energy-momentum tensor is defined as below~\cite{HILGEVOORD19631,HILGEVOORD19651002} 
 \begin{align}
\hat{T}^{\mu\nu}_{\text{HW}}= \hat{T}^{\mu\nu}_{\text{Can}} +\frac{i}{2m} \left(\partial^{\nu}\psbar \sigma^{\mu\beta}\partial_\beta \psi +\partial_\alpha \psbar \sigma^{\alpha\mu}\partial^{\nu}\psi\right)- \frac{i}{4m}g^{\mu\nu} \partial_\lambda \left(\psbar\sigma^{\lambda\alpha}\overleftrightarrow{\partial}_\alpha \psi\right),
\end{align}
with $\sigma_{\mu\nu} \equiv (i/2) \left[\gamma_\mu,\gamma_\nu \right]$. Here the thermal average and fluctuation are calculated respectively as
\begin{align}
&\langle:\hat{T}^{00}_{\text{HW},a}:\rangle
= 4\int\frac{d^3k}{(2\pi)^3}~\omega_{\boldsymbol{k}}f(\omega_{\boldsymbol{k}}) \equiv\varepsilon_{\text{HW}}(T) 
\label{equ18ver1}\\
&\sigma^2_{\text{HW}}(a,m,T) =  \frac{2}{m^2}\int dK dK^{\prime} f(\omega_{{\boldsymbol{k}}}) \Big[\left(\omega_{{\boldsymbol{k}}}\omega_{{\boldsymbol{k}}^{\prime}}+{\boldsymbol{k}}\cdot{\boldsymbol{k}}^{\prime}+m^2\right)^2\left(\omega_{{\boldsymbol{k}}}\omega_{{\boldsymbol{k}}^{\prime}}-{\boldsymbol{k}}\cdot{\boldsymbol{k}}^{\prime}+m^2\right)\nonumber\\
& \times e^{-\frac{a^2}{2}({\boldsymbol{k}}-{\boldsymbol{k}}^{\prime})^2}-(\omega_{{\boldsymbol{k}}}\omega_{{\boldsymbol{k}}^{\prime}}+{\boldsymbol{k}}\cdot{\boldsymbol{k}}^{\prime}-m^2)^2(\omega_{{\boldsymbol{k}}}\omega_{{\boldsymbol{k}}^{\prime}}-{\boldsymbol{k}}\cdot{\boldsymbol{k}}^{\prime}-m^2)\times e^{-\frac{a^2}{2}({\boldsymbol{k}}+{\boldsymbol{k}}^{\prime})^2}\Big](1-f(\omega_{{\boldsymbol{k}}^{\prime}})).
\label{equ19ver1}
\end{align}
It can be seen easily from Eqs.~\eqref{equ11ver1}, \eqref{equ15ver1}, and \eqref{equ18ver1} that, $\varepsilon_{\text{Can}}(T)=\varepsilon_{\text{GLW}}(T)=\varepsilon_{\text{HW}}(T)$, while the fluctuations of $:\hat{T}^{00}_a:$ are different for different choice of pseudo-gauge. 
Eqs.~\eqref{equ12ver1}, \eqref{equ16ver1}, and \eqref{equ19ver1} are used to calculate the fluctuations of energy-density of the subsystem $S_a$ of the larger system $S_V$.
Both the energy density ($\varepsilon$) and fluctuation ($\sigma$) can be extended to incorporate other degeneracy factors such as  isospin or color charge degrees of freedom.
\section{Summary}
We have calculated the mathematical expressions for quantum energy-density fluctuations for the subsystems of hot relativistic gas of spin-$\frac{1}{2}$ particles. Our results show that even though the energy-density for all choices of pseudo-gauge are same, still their fluctuations depend on the forms of pseudo-gauge, which means that it is pseudo-gauge dependent~\cite{Mrowczynski:1997mj,Becattini:2012pp,Nakayama:2012vs} and should be kept in mind during the experimental measurements.

I am grateful to A. Das, W. Florkowski, and R. Ryblewski for their intriguing collaboration. This research was supported in part by the Polish National Science Centre Grants No. 2016/23/B/ST2/00717 and No. 2018/30/E/ST2/00432, and IFJ PAN.

\end{document}